# Nuclear equation of state and incompressibility in a model with correlations from giant monopole vibrations

Aram Z. Mekjian and Larry Zamick

Department of Physics and Astronomy, Rutgers University, Piscataway, NJ 08854

Abstract

The nuclear equation of state and incompressibility are studied in a model with correlations. We first consider a Hartree solution and then add one-particle-one hole (1p-1h) and two particle-two hole (2p-2h) admixtures into the wavefunction using a separable monopole-monopole interaction $\lambda r^2(i) r^2(j)$ which induces giant monopole vibrations. Stable and unstable point arise in the nuclear equation of state as a function of the oscillator size parameter from pure 1p-1h admixtures. Addition of 2p-2h admixtures lead to one stable point. However, the incompressibility is found to be quite different from the Hartree case with 1p-1h admixtures, even with 2p-2h states included.

1. Introduction

In a previous publication by Zamick [1] a separable monopole-monopole interaction was used to evaluate the nuclear incompressibility in a Hartree approximation. In this work we study the effects of correlations using the same interaction. The incompressibility $\kappa$ and its associated equation of state (EOS) is a quantity of importance in heavy ion collisions and supernovae explosions. There is some uncertainty in the value of $\kappa$ in both non-relativistic and relativistic mean field approaches. Non-relativistic models [2,3,4] have values of $\kappa \approx 220 - 233 MeV$ while relativistic models [5], which include contributions from the negative energy sea, have somewhat higher values with $\kappa \approx 250 - 270 MeV$. The experimental determination of $\kappa$ comes from a study of the isoscalar giant monople resonance or the breathing mode state. The incompressibility in asymmetric nuclear matter is also of recent interest [6-10]. The effect of pairing correlations on the incompressibility in nuclear matter was studied in ref.[11]. Here we study the role of monopole correlations on the equation of state and its associated incompressibility.

2. Hartree Model

The incompressibity $\kappa$ is defined by second derivatives of the energy per particle $E/A$ with respect the total radius $R$ or volume $V$ as:

$$\kappa = R^2 \frac{d^2(E/A)}{dR^2} = 9V^2 \frac{d^2(E/A)}{dV^2} \qquad (1)$$

The $\kappa$ is evaluated at the minimum of $E/A$ with varying $R$. The calculations presented will be for a harmonic oscillator. A monopole-monopole interaction $\lambda r^2(i) r^2(j)$ acting between nucleons leads to a one-body harmonic oscillator potential in a Hartree approximation. Particles can then be put into the lowest levels of this one-body potential to form a state given by a Slater determinate. The same interaction can then be used to study the role of correlations, both 2p-2h and 1p-1h states. One of us studied features associated with this model in an early work [1]. The present paper extends some the results to include the role of correlations on the nuclear incompressibility. The importance of correlations was also investigated in Ref.[12,13] from another perspective. The single particle orbits of a harmonic oscillator potential with quantum numbers $n,l$ in polar coordinates have energies given by $\varepsilon_{n,l} = (2n + l + 3/2)\hbar\omega$. In Cartesian coordinates this reads $\varepsilon_N = (N + 3/2)\hbar\omega$ with $N = n_x + n_y + n_z$. The degeneracy, including spin degeneracy $g_s = 2$, is $(N+1)(N+2)$. An important connection for a three dimensional harmonic oscillator is

$$A<r^2> = \frac{3}{5}A^{5/3}r_0^2 = \frac{\hbar}{M\omega_0}2\sum_{N=0}^{N_{max}}(N+\frac{3}{2})(N+1)(N+2) \approx \frac{\hbar}{2M\omega_0}(N_{max}+2)^4 \quad (2)$$

The $N_{max}$ is the $N$ of the last filled shell. This result will be useful for obtaining the behavior of the incompressibility with mass number $A$. Let $\Sigma$ denote the sum of $\varepsilon_{n,l}/\hbar\omega$ over all occupied states. For example for $^{16}O$, $\Sigma = 36$. The oscillator frequency $\omega$ and oscillator size parameter $b$ are connected by $\hbar\omega = \hbar^2/mb^2$. The $b$ determines the radius through $A<r^2> = \Sigma \cdot b^2 = (3/5)r_0^2 A^{2/3}$. For the equilibrium position of $b = b_0$ the $\hbar\omega_0 = \hbar^2/mb_0^2$ and $A$ are connected by the well known relation $\hbar\omega_0 = 41 MeV/A^{1/3}$ and the $r_0 = 1.2 fm$.

Varying the size parameter $b$ or radius $R$ corresponds to varying $\omega$. The $\hbar\omega = \hbar^2/mb^2$ can therefore be regarded as a variational parameter. The Hartree energy of the separable monopole-monopole model is [1]

$$E(\hbar\omega) = \frac{\hbar\omega}{2}\Sigma + \frac{\tilde{\lambda}}{2(\hbar\omega)^2}\Sigma^2 \quad (3)$$

or

$$E(b) = \frac{\hbar^2}{2mb^2}\Sigma + \frac{\tilde{\lambda}m^2b^4}{2\hbar^4}\Sigma^2 \quad (4)$$

The first term is the kinetic energy contribution to $E(\hbar\omega)$ while the second term arises from the separable monopole-monopole interaction. The $\Sigma^2/(\hbar\omega)^2$ factor reflects the behavior of $<r^2>$ for each separate monopole term: $\Sigma_i m\omega^2 <r(i)^2> \sim \hbar\omega\Sigma$. The $\tilde{\lambda} = \lambda(\hbar^2/m)^2$. The $E(\hbar\omega)$ has a minimum with respect to variation with $\hbar\omega$ at $(\hbar\omega_0)^3 = 2\tilde{\lambda}\Sigma$. The incompressibility involves the second derivative with respect to $R$ or $b$ with $\kappa = b^2 d^2(E/A)/db^2$ evaluated at $b_0$ where $\hbar\omega_0 = \hbar^2/mb_0^2$. The resulting $\kappa$ is

$$\kappa = 6\hbar\omega_0 \Sigma / A. \tag{5}$$

A numerical value can be obtained from $\hbar\omega_0 = 41 MeV / A^{1/3}$ and $\Sigma / A \approx (3/5)r_0^2 A^{1/3}$ giving $\kappa = 6 \cdot 41 \cdot 0.6 \cdot (1.2)(1.2) MeV = 213 MeV$. The result of $\kappa = 213 MeV$ is near the range of values $\kappa \approx 220 - 233 MeV$ noted above.

3. Role of correlations

We now proceed to evaluate the role of correlation on the energy functional and the incompressibility. As mentioned above, the separable monopole-monopole interaction can then be used to study the role of correlations, both 2p-2h and 1p-1h states. The admixtures plus the 0p-0h ground state are written as

$$|\psi\rangle = \alpha|0p-0h\rangle + \beta|1p-1h\rangle + \gamma|2p-2h\rangle \tag{6}$$

The normalization constraint is $\alpha^2 + \beta^2 + \gamma^2 = 1$. The energy associated with this state is developed in ref.[1]

$$\frac{E(x)}{\hbar\omega_0} = \frac{\Sigma}{2x} + \frac{x^2\Sigma}{4} + 2\alpha(\beta,\gamma)\frac{\beta\sqrt{\Sigma}}{2}(x^2 - \frac{1}{x}) + \beta^2(\frac{3x^2}{2} + \frac{1}{x}) + 2\alpha(\beta,\gamma)\frac{\sqrt{2}\gamma x^2}{2}$$
$$+ 2\gamma^2(\frac{3x^2}{2} + \frac{1}{x}) + 2\beta\gamma\frac{\sqrt{2\Sigma}}{2}(x^2 - \frac{1}{x}) \tag{7}$$

with $\alpha(\beta,\gamma) = \sqrt{1-\beta^2-\gamma^2}$ from the constraint. Here the variable $x = \hbar\omega_0 / \hbar\omega$. For $^{16}O$, the lowest energy is obtained for $\beta = -0.06, \gamma = -0.138, \alpha(\beta,\gamma) = 0.998861$ and $x = 1.022$ giving $E(x)/\hbar\omega_0 = 26.8986$. The pure 0p-0h has $\beta = 0, \gamma = 0, \alpha(\beta,\gamma) = 1$ and has $x = 1$ with an associated $E(x)/\hbar\omega_0 = 27$. The energy in units of $\hbar\omega_0$ for the 1p-1h and 2p-2h states depends on the dimensionless variable $x = \hbar\omega_0 / \hbar\omega$. We take the 1p-1h state at $(3x^2/2 + 1/x)$ and the 2p-2h state at twice this energy. In perturbation theory the amplitude $\beta = <0|H|r^2> /(\Delta E\sqrt{<r^2/r^2>})$ is then

$$\beta = -\frac{1}{2}\sqrt{\Sigma}\frac{x^2 - 1/x}{((3/2)x^2 + 1/x)} \tag{8}$$

Similarly, the amplitude $\gamma = <0|\lambda r^2 r^2|r^2 r^2> /(\Delta E\sqrt{<r^2|r^2>})$ is

$$\gamma = -\frac{\sqrt{2}}{4}\frac{x^2}{((3/2)x^2 + 1/x)} \tag{9}$$

Substituting these expressions into the equation for $E/\hbar\omega_0 \equiv f(\Sigma,x)$ gives

$$\frac{E}{\hbar\omega_0} \equiv f(\Sigma,x) = \frac{\Sigma}{2x} + \frac{x^2\Sigma}{4} - \frac{1}{2}(\alpha-\frac{1}{2})\Sigma x \frac{(x^2-1/x)^2}{((3/2)x^2+1/x)} - \frac{1}{2}(\alpha-\frac{1}{2})\frac{x^4}{((3/2)x^2+1/x)} +$$

$$\frac{\Sigma}{4}\frac{x^2(x^2-1/x)^2}{((3/2)x^2+1/x)^2} \tag{10}$$

We define $g(\Sigma,x)$ to be the first two terms of $f(\Sigma,x)$ or

$$g(\Sigma,x) = \frac{\Sigma}{2x} + \frac{x^2\Sigma}{4} \tag{11}$$

which is the uncorrelated $E/\hbar\omega_0$. In Fig.1 we plot $f(\Sigma,x)$ & $g(\Sigma,x)$ and in Table1 we give values of $f''(\Sigma,x_{min})$ and $g''(\Sigma,x_{min})$ and the minimum $x$ for a large range of $\Sigma$.

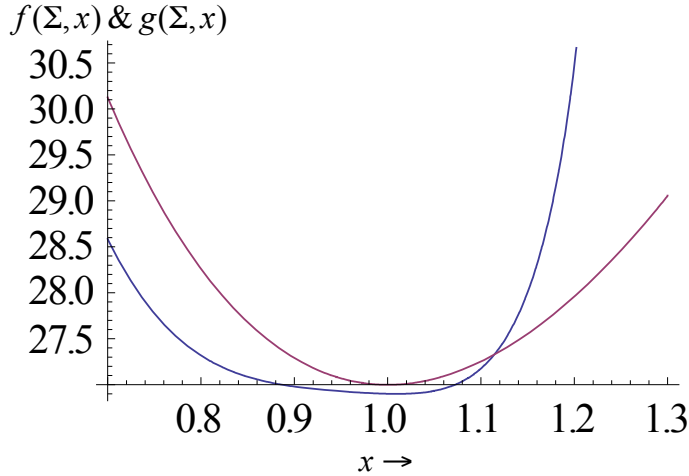

FIG.1. (Color online). The behavior of $f(\Sigma,x)$ & $g(\Sigma,x)$ as a function of the oscillator parameter $x$ for $\Sigma=36$. The flatter curve is $f(\Sigma,x)$.

Table 1. Behavior of $f(\Sigma,x)$ and $g(\Sigma,x)$ at the minimum value of $x$ for various $\Sigma$. For large values of $\Sigma$, the ratio of $f''(\Sigma,x_{min})/g''(\Sigma,x_{min}) \approx 0.35$. The $g''(\Sigma,x_{min})= 3\Sigma/2$. Thus, for large $\Sigma$, the $f''(\Sigma,x_{min})$ scales with $\Sigma$ and is $f''(\Sigma,x_{min}) \approx .35(3\Sigma/2)$. For large $\Sigma$, $f(\Sigma,x_{min})=g(\Sigma,x=1)=3\Sigma/4$. Even at low $\Sigma$, $f(\Sigma,x_{min}) \approx g(\Sigma,x=1)$ as can be seen from Fig.1 for $\Sigma=36$. Since $x=b^2/b_0^2$ the minimum of $f(\Sigma,b)$ and $g(\Sigma,b)$ occur $b/b_0 = \sqrt{x_{min}}$ and the $f''(\Sigma,b)$ and $g''(\Sigma,b)$ are 4 times the corresponding

values given in the two columns.

| $\Sigma$ | $x = x_{min}$ | $f''(\Sigma, x_{min})$ | $x = x_{min}$ | $g''(\Sigma, x_{min})$ |
|---|---|---|---|---|
| 36 | 1.013 | 26.8 | 1. | 54 |
| 72 | 1.0071 | 46.2 | 1. | 108 |
| 108 | 1.0049 | 65.4 | 1. | 162 |
| 208 | 1.0026 | 118.4 | 1. | 312 |
| 312 | 1.0018 | 173.4 | 1. | 468 |
| 416 | 1.0013 | 228.5 | 1. | 624 |
| 520 | 1.00011 | 283.5 | 1. | 780 |
| 624 | 1.00089 | 338.5 | 1. | 936 |
| 728 | 1.00077 | 393.6 | 1. | 1092 |
| 832 | 1.00067 | 448.6 | 1. | 1248 |
| 936 | 1.00060 | 503.6 | 1. | 1404 |
| 1040 | 1.00054 | 558.6 | 1. | 1560 |

4. Correlation results from matrix diagonalization

4.1. Correlations in the complete 0p-0h, 1p-1h and 2p-2h admixtures; the $\alpha, \beta, \gamma$ model.

First we see large differences between the correlated and Hartree model. Secondly, we would like to study large ranges of values of $x$ were perturbation theory could have important corrections. We therefore proceed by studying the role of correlations in our model by exact diagonalizations of the coupling matrix between several situations which first include both 1p-1h and 2p-2h components, and then study the role of each component separately. In this way we can see which couplings are responsible for any large departures.

$$\begin{pmatrix} \frac{\Sigma}{2x} + \frac{\Sigma x^2}{4}, \frac{\sqrt{\Sigma}}{2}(x^2 - \frac{1}{x}), \frac{x^2}{\sqrt{2}} \\ \frac{\sqrt{\Sigma}}{2}(x^2 - \frac{1}{x}), \frac{\Sigma}{2x} + \frac{\Sigma x^2}{4} + \frac{3x^2}{2} + \frac{1}{x}, \sqrt{2\Sigma}(\frac{x^2}{2} - \frac{1}{2x}) \\ \frac{x^2}{\sqrt{2}}, \sqrt{2\Sigma}(\frac{x^2}{2} - \frac{1}{2x}), \frac{\Sigma}{2x} + \frac{\Sigma x^2}{4} + 3x^2 + \frac{2}{x} \end{pmatrix} \quad (12)$$

The eigenvalues of this matrix are plotted in Fig.2. The behavior of the lowest eigenvalue is also shown

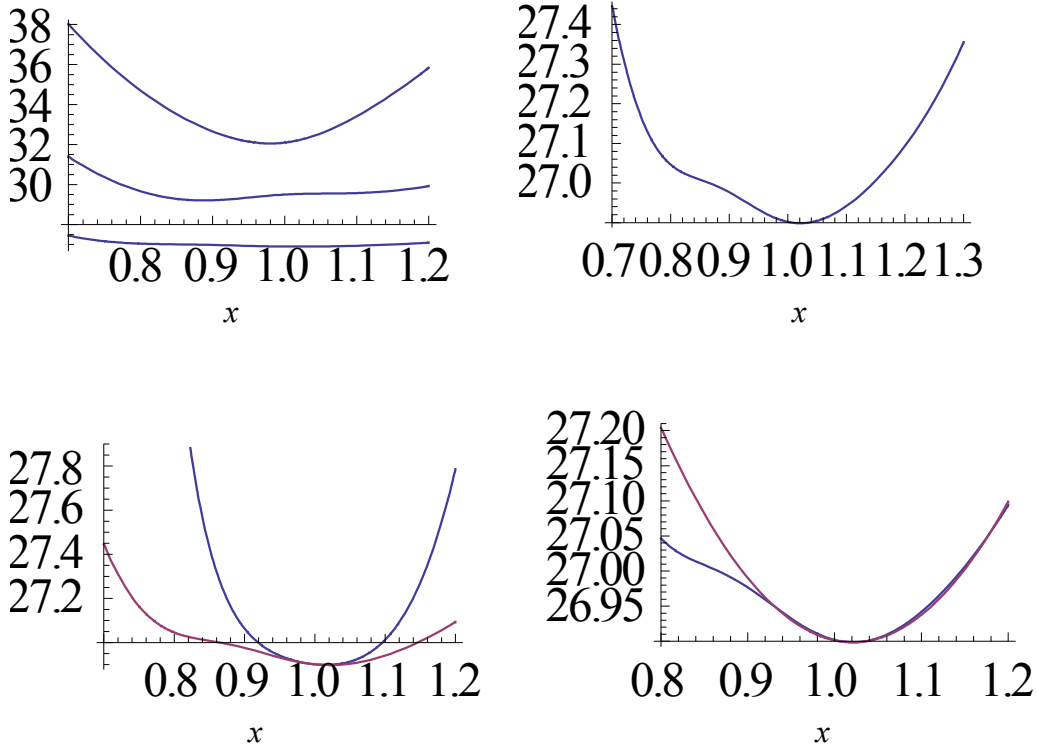

FIG.2. (Color online) The eigenvalues of the 3x3 coupling matrix. The upper left figure shows the behavior of the three eigenvalues with variation of the parameter $x$. The upper right figure shows the behavior of the lowest eigenvalues with variation of the parameter $x$. The lower left figure compares the perturbation result and the lowest eigenvalue curve. The inner curve is the perturbation result. The lower right figures is a fit to the lowest eigenvalue curve using $12.5(x-1.021)^2/2 + 26.9$. The 12.5 gives the approximate curvature at the minimum where the minimum has a value 26.9 at $x=1.021$.

4.2. Correlations limited to 0p-0h plus 1p-1h admixtures; the $\alpha, \beta$ model.
We next consider the case of 1p-1h admixtures to the 0p-0h ground state. The 2x2 coupling matrix is simply the upper 2X2 part of the 3X3 matrix; namely:

$$\begin{pmatrix} \frac{\Sigma}{2x} + \frac{\Sigma x^2}{4}, & \frac{\sqrt{\Sigma}}{2}(x^2 - \frac{1}{x}) \\ \frac{\sqrt{\Sigma}}{2}(x^2 - \frac{1}{x}), & \frac{\Sigma}{2x} + \frac{\Sigma x^2}{4} + \frac{3x^2}{2} + \frac{1}{x} \end{pmatrix} \quad (13)$$

The eigenvalues of this matrix are easily found and are

$$E_{\pm}^{(1)}(\Sigma,x) = \frac{2(1+\Sigma)+(3+\Sigma)x^3 \pm \sqrt{4\Sigma(x^3-1)^2+(2+3x^3)^2}}{4x} \qquad (14)$$

For $\Sigma = 36$, the unnormalized eigenvectors are

$$V_{\pm}^{(1)}(36,x) = \left\langle \frac{2+3x^3 \pm \sqrt{148-276x^3+153x^6}}{12(1-x^3)}, 1 \right\rangle \qquad (15)$$

Fig.3 shows a plot of the lowest eigenvalue $E_{-}^{(1)}(\Sigma,x)$ versus $x$ for $\Sigma = 36$ and compares it to the lowest eigenvalue of the complete 3X3 matrix. We note the following features for the lowest eigenvalue as a function of $x$. The $E_{-}^{(1)}(\Sigma,x)$ has a two minima and a local maxima. The maxima occurs at $x = 1$, while the minima occur on either side of the maxima at $x = 0.912$ and $x = 1.092$. The smaller $x$ minima is slightly deeper. A comparison with the lowest eigenvalue including both 1p-1h and 2p-2h admixtures is shown in the lower part of this figure. The shoulder in the 3X3 case has its origin in the first minima of the 1p-1h admixture. The minima of the 3X3 case is at $x=1.021$ while the second minima of the 1p-1h admixture case is at a higher value, namely $x = 1.092$.

$$E_{-}^{(1)}(36,x)$$

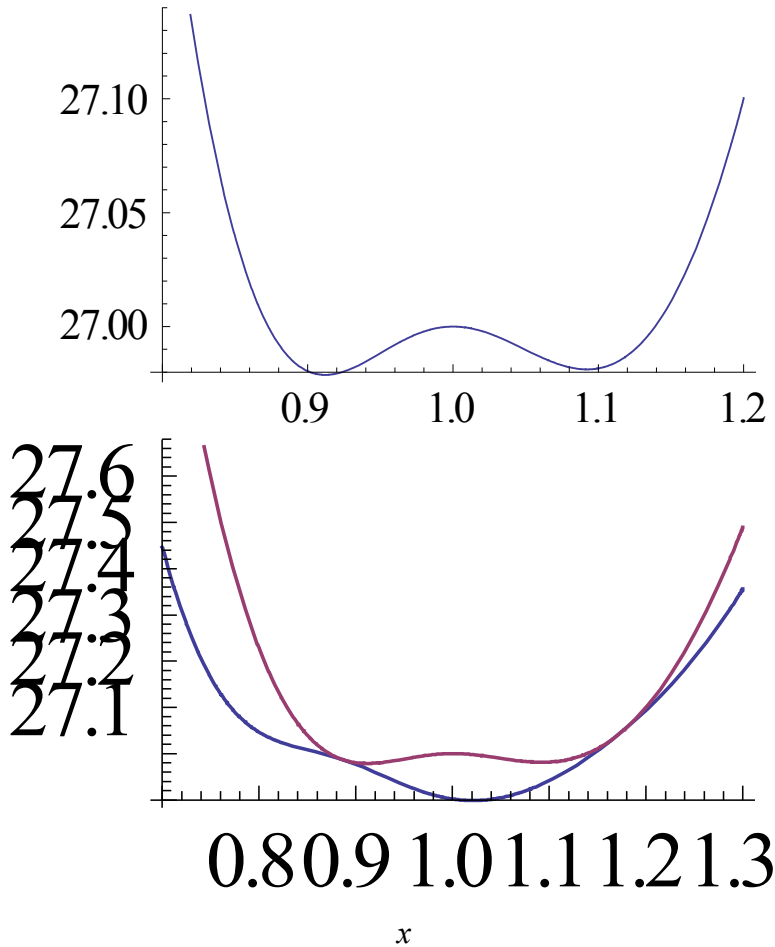

FIG.3. (Color online). The behavior of the lowest eigenvalue with oscillator parameter $x$. The upper figure is an $\alpha, \beta$ model with only 1p-1h admixtures. The lower figure is a comparison with the case of adding 2p-2h states, or $\alpha, \beta, \gamma$ model, into the description. The point $x=1$, which was a stable point (minima with positive curvature) in the Hartee case, is now an unstable point (maxima with negative curvature) in the $\alpha, \beta$ model. The maximum point is surrounded on the left and right side by two minima almost equally spaced from the center $x=1$ point. The depth of the minima are almost the same, with the right minima slightly lower in energy. From the lower figure we see that the 2p-2h admixture leads to a single minima displaced from $x=1$.

4.3. Correlations limited to 0p-0h plus 2p-2h admixtures; the $\alpha, \gamma$ model.

The case of 0p-0h plus 2p-2h states is obtained by diagonalizing a 2X2 matrix which contain the elements 11, 13, 31, 33 of the 3X3 matrix. The two eigenvalues, called $E^{(2)}_{\pm}(\Sigma, x)$, are

$$E^{(2)}_{\pm}(\Sigma, x) = \frac{(2+\Sigma) + (3+\Sigma/2)x^3 \pm \sqrt{4 + 12x^3 + 11x^6}}{2x} \tag{16}$$

The behavior of the lowest eigenvalue is shown in Fig.4. The left figure in Fig.4 shows a plot of $E^{(2)}_{-}(\Sigma, x)$ versus $x$ for $\Sigma = 36$ along with the Hartree model. The right figure in Fig.4 shows the comparison of $E^{(1)}_{-}(\Sigma, x)$, $E^{(2)}_{-}(\Sigma, x)$ and the lowest eigenvalue of the complete 3X3 matrix. From the figures we see that the 2p-2h admixture alone makes only a small difference with respect to the uncorrelated Hartree case. A comparison of $E^{(1)}_{-}(36, x)$ and $E^{(2)}_{-}(36, x)$ shows that the 1p-1h admixture produces large departures from the 2p-2h case and, in turn, from the Hartree case which is very close to $E^{(2)}_{-}(36, x)$.

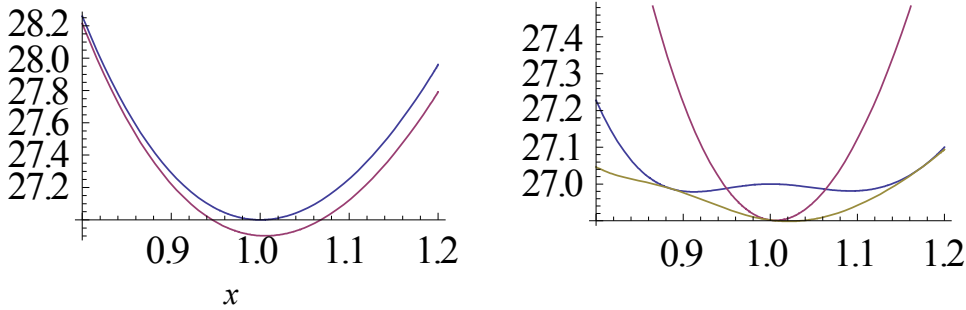

FIG. 4. (Color online). Lowest eigenvalue $E^{(2)}_{-}(36, x)$ of the 2p-2h model. The left figure compares $E^{(2)}_{-}(36, x)$ with the Hartree model. The upper curve is the Hartree model. The minimum of $E^{(2)}_{-}(36, x)$ is slightly shifted from the minimum of the Hartree model of $x=1$ to $x=1.0058$. The right figure compares $E^{(1)}_{-}(36, x)$, $E^{(2)}_{-}(36, x)$ and the lowest eigenvalue of the complete 3X3 matrix. The $E^{(1)}_{-}(36, x)$ curve has the maxima and two minima as already discussed. The $E^{(2)}_{-}(36, x)$ is the more parabolic inner curve. The lowest curve is the lowest eigenvalue curve.

5. Nuclear incompressibility

We now turn our attention to the incompressibility and how correlations influence ite value. To obtain the incompressibility it is necessary to evaluate the second derivative in Eq.1 with respect to variations in the variable $R$. The results so far give behaviors with respect to $x$. To obtain the appropriate variation we proceed as follows. The incompressibility follows from

$$\kappa = <r^2> \frac{d^2 E/A}{d<r^2>^{1/2} d<r^2>^{1/2}} \tag{17}$$

when evaluated at the minimum $<r^2>^{1/2}$. The $<r^2> \equiv <r^2>_C$ is given by

$$\frac{<r^2>_C}{<r^2>_H} \equiv <r^2>_R = \frac{x}{\Sigma}(\Sigma + 2\beta^2 + 4\gamma^2 + 2\alpha\beta\sqrt{\Sigma} + 2\beta\gamma\sqrt{\Sigma}) \tag{18}$$

where $<r^2>_C$ is the correlated result and $<r^2>_H = \Sigma b^2/A = (3/5)r_0^2 A^{2/3}$ is the uncorrelated Hartree result.

The incompressibility can be written as

$$\kappa = <r^2>_R \frac{d^2 E/A}{d<r^2>_R^{1/2} d<r^2>_R^{1/2}} = <r^2>_R (\frac{dx}{d\sqrt{<r^2>_R}})^2 \frac{d^2(E/A)}{(dx)^2}$$

$$= 4<r^2>_R^2 (\frac{1}{d<r^2>_R/dx})^2 \frac{d^2(E/A)}{(dx)^2} \tag{19}$$

A linear term does not contribute since $\kappa$ is evaluated at the minimum. The minimum is determined by

$$\frac{d(E/A)}{d<r^2>_R^{1/2}} = (\frac{dx}{d\sqrt{<r^2>_R}})\frac{d(E/A)}{(dx)} = 0 \tag{20}$$

which occurs at $d(E/A)/dx = 0$. The $\alpha, \beta, \gamma$ that are necessary are elements of the eigenvectors, and in particular, we want the eigenvector for the lowest eigenvalue. Since the 1p-1h admixture makes a significant change in the behavior of the energy, we begin by studying this simpler case first which has $\alpha, \beta$ only. The un-normalized eigenvector is given by Eq.15. The curvature $d^2(E/\hbar\omega_0)/(dx)^2 = 14$ at the right-most minimum at $x = 1.09221$ or $<r^2>_R^{1/2} = 1.00066$. The $d<r^2>_R^{1/2}/dx = 0.14$. Consequently,

$$<r^2>_R d^2(E/\hbar\omega_0)/(d\sqrt{<r^2>_R})^2 = 14/.14^2 \sim 700. \tag{21}$$

The Hartree model has $d^2(E/\hbar\omega_0)/(d\sqrt{<r^2>_R})^2 = 54\text{x}4$. The ratio of the correlated to Hartree model for this curvature is then 700/(216)~3.24. This ratio also reflects the ratio of incompressibilities.

The full $\alpha, \beta, \gamma$ model has $d^2(E/\hbar\omega_0)/(dx)^2 = 12.5$ and $d<r^2>_R^{1/2}/dx = 0.1422$ and thus

$$<r^2>_R d^2(E/\hbar\omega_0)/(d\sqrt{<r^2>_R})^2 = 12.5/.142^2 \sim 620 \tag{22}$$

The value 620 is 2.87 times the Hartree model result. As noted above, this ratio also reflects the ratio of incompressibilities.

5. Conclusions

In conclusion we studied the behavior of the energy functional in a Hartree model and a correlated model. An a priori simple problem took some strange turns when correlations were included. The correlations were induced using a monopole interaction. The role of 2p-2h and 1p-1h admixtures were included. The 2p-2h admixture alone did not significantly change the behavior of the energy as a function of the oscillator parameter $x$ or reduced radius $<r^2>_R^{1/2}$ from the Hartree model. By contrast, the 1p-1h admixture lead to significant changes in both the the energy as a function of the oscillator parameter $x$ or reduced radius $<r^2>_R^{1/2}$. For example, the point $x=1$, which was a stable point (minima with positive curvature) in the Hartee case, is now an unstable point (maxima with negative curvature) in the $\alpha, \beta$ model. The maximum point is surrounded on the left and right side by two minima almost equally spaced from the center point. When 2p-2h admixtures are included their result leads to a single minima displaced from $x=1$. Also, the curves as a function of the oscillator size parameter $x$ for the admixture of both 1p-1h and 2p-2h correlations are very flat around the equilibrium point compared to the near parabolic shape of the Hartree or Hartee plus 2p-2h model. This feature might suggest a low incompressibility. However, we found that the 1p-1h correlations increase the incompressibility by a factor of about 3 from the uncorrelated case for this particular interaction. The source of this higher incompressibility lies in the behavior of the energy with radius. Specifically, the incompressibility involves the second derivative with respect to the radius. The second derivative with respect to radius is enhanced from the second derivative with respect to the oscillator parameter by the square of the rate of change of radius $x$ with respect to radius which turns out to be large. The present work suggests further studies of the interplay of 1p-1h and 2p-2h admixtures with realistic interactions and their effect on the incompressibility.

Acknowledgements. Supported in part by a DOE grant DE-FG02-96ER-4097.

References.
1. L.Zamick, Nucl. Phys. A232, 13 (1974)
2. J.Piekarewicz, Phys. Rev. C69, 041301 (2004)
3  V.B.Soubbotin, V.I.Tselyaev,X.Vinas, Phys. Rev. C69, 064312 (2004)
4. D.Vretenar, T.Niksic, P.Ring, Phys. Rev.  C68, 024319 (2003)
5. G.A.Lalazissis, J.Konig, P.Ring, Phys. Rev. C55, 540  (1997)
6. Lie-Wen Chen, Bao-Jun Cai, Chun Shen, Che Ming Ko, Jun Xu, Bao-An Li arXiv:0910.0086
7. D.N. Basu, P. Roy Chowdhury, C. Samanta, Phys. Rev.C80:057304 (2009)
8. E.Khan, Phys.Rev.C80:057302 (2009)
9. J. M. Pearson, N. Chamel, S. Goriely, Phys.Rev.C82:037301 (2010)
10. J.Piekarewicz, M.Centelles  Phys.Rev.C79:054311 (2009)
11. E. Khan, J. Margueron, G. Colo, K. Hagino, H. Sagawa, arXiv:1005.1741
12. A.Z.Mekjian, S.J.Lee, L.Zamick, Phys. Rev. C72 (2005) 044305
13. A.Z.Mekjian, S.J.Lee, L.Zamick, Phys. Lett. B621 (2005) 239-245